
\input phyzzx

%
\catcode`\@=11 
\def\papers{\papersize\headline=\paperheadline\footline=\paperfootline}
\def\papersize{\hsize=40pc \vsize=53pc \hoffset=0pc \voffset=1pc
   \advance\hoffset by\HOFFSET \advance\voffset by\VOFFSET
   \pagebottomfiller=0pc
   \skip\footins=\bigskipamount \normalspace }
\catcode`\@=12 
\papers

\def\to{\rightarrow}
\def\half{\textstyle{1\over 2}}

\vsize=21.5cm
\hsize=15.cm

\tolerance=500000
\overfullrule=0pt

\pubnum={PUPT-1373 \cr
hep-th@xxx/9301021\cr
January 1993,\cr
revised and extended\cr
April 1993}

\date={}
\pubtype={}
\titlepage
\title{POSITIVE ENERGY THEOREM AND SUPERSYMMETRY
IN \break
EXACTLY SOLVABLE QUANTUM-CORRECTED\break
2D DILATON-GRAVITY}
\author{{
Adel~Bilal}\foot{
 on leave of absence from
Laboratoire de Physique Th\'eorique de l'Ecole
Normale Sup\'erieure, \nextline 24 rue Lhomond, 75231
Paris Cedex 05, France
(unit\'e propre du CNRS)\nextline
e-mail: bilal@puhep1.princeton.edu
}}
\address{\it Joseph Henry Laboratories\break
Princeton University\break
Princeton, NJ
08544, USA}

\vskip 3.mm
\abstract{Extending the work of Park
and Strominger,  we prove a positive energy theorem
for the exactly solvable quantum-corrected
2D dilaton-gravity theories. The positive energy functional
we construct is shown to be unique (within a reasonably
broad class of such functionals). For field configurations
asymptotic to the LDV we show that this energy functional
(if defined on a space-like surface) yields the usual
(classical) definition of the ADM  mass {\it plus} a
certain ``quantum"-correction. If defined on a null
surface the energy functional yields the Bondi-mass. The
latter is evaluated carefully and applied to the RST
shock-wave scenario where it is shown to behave as
physically expected.
Motivated by the existence of a positivity theorem we
construct manifestly supersymmetric (semiclassical)
extensions of these quantum-corrected dilaton-gravity
theories. }

\endpage
\pagenumber=1

 \def\PL #1 #2 #3 {Phys.~Lett.~{\bf #1} (#2) #3}
 \def\NP #1 #2 #3 {Nucl.~Phys.~{\bf #1} (#2) #3}
 \def\PR #1 #2 #3 {Phys.~Rev.~{\bf #1} (#2) #3}
 \def\PRL #1 #2 #3 {Phys.~Rev.~Lett.~{\bf #1} (#2) #3}
 \def\CMP #1 #2 #3 {Comm.~Math.~Phys.~{\bf #1} (#2) #3}
 \def\IJMP #1 #2 #3 {Int.~J.~Mod.~Phys.~{\bf #1} (#2) #3}
 \def\JETP #1 #2 #3 {Sov.~Phys.~JETP.~{\bf #1} (#2) #3}
 \def\PRS #1 #2 #3 {Proc.~Roy.~Soc.~{\bf #1} (#2) #3}
 \def\IM #1 #2 #3 {Inv.~Math.~{\bf #1} (#2) #3}
 \def\JFA #1 #2 #3 {J.~Funkt.~Anal.~{\bf #1} (#2) #3}
 \def\LMP #1 #2 #3 {Lett.~Math.~Phys.~{\bf #1} (#2) #3}
 \def\IJMP #1 #2 #3 {Int.~J.~Mod.~Phys.~{\bf #1} (#2) #3}
 \def\FAA #1 #2 #3 {Funct.~Anal.~Appl.~{\bf #1} (#2) #3}
 \def\AP #1 #2 #3 {Ann.~Phys.~{\bf #1} (#2) #3}
 \def\MPL #1 #2 #3 {Mod.~Phys.~Lett.~{\bf #1} (#2) #3}

\def\d{\partial}
\def\dpl{\partial_+}
\def\dm{\partial_-}
\def\Scl{S_{\rm cl}}
\def\Sanom{S_{\rm anom}}

\def\f{\phi}
\def\r{\rho}
\def\o{\omega}
\def\O{\Omega}
\def\x{\chi}
\def\s{\sigma}
\def\sp{\sigma^+}
\def\sps{\sigma^+_s}
\def\spt{{\tilde\sigma^+}}
\def\spo{\sigma^+_0}
\def\so{\sigma_0}
\def\sm{\sigma^-}
\def\sms{\sigma^-_s}
\def\smt{\tilde\sigma^-}
\def\l{\lambda}
\def\t{\tau}
\def\ix{\int {\rm d}^2 x}
\def\ixt{\int {\rm d}^2 x {\rm d}^2 \theta}
\def\is{\int {\rm d}^2 \sigma}
\def\rg{\sqrt{-g}}
\def\k{\kappa}
\def\eom{equation of motion\ }
\def\eoms{equations of motion\ }
\def\m{\mu}
\def\n{\nu}
\def\tmn{T_{\mu\nu}}
\def\N{\nabla}
\def\zt{{\tilde Z}}
\def\S{\Sigma}
\def\e{\epsilon}
\def\eb{\bar\epsilon}
\def\a{\alpha}
\def\g{\gamma}
\def\gf{\gamma_5}
\def\G{\Gamma}
\def\NS{ {/\kern -0.60em \nabla } }
\def\M{{\cal M}}
\def\Or{{\cal O}}
\def\kk{{\k\over 4}}
\def\mb{M_{\rm B}}

{ \chapter{Introduction}}

Dilaton-gravity in two dimensions provides a simplified
model to study quantum gravity and in particular the
analogues of (four-dimensional) black hole formation and
evaporation. The classical action for dilaton-gravity
coupled to $N$ conformal massless matter fields
\REF\CGHS{C. Callan, S. Giddings, J. Harvey and A.
Strominger, \PR D45 1992 R1005 .} [\CGHS]
$$
\Scl = {1\over 2\pi} \ix \rg \left[ e^{-2\f}\left(
R+4(\nabla\f)^2 + 4\l^2\right) -\half \sum_{i=1}^N
(\nabla f_i)^2\right]
\eqn\i$$
admits classical (non-radiating) black hole solutions
$$
{\rm d}s^2 = -{{\rm d}x^+ {\rm d}x^-\over
{m\over \l} -\l^2 x^+ x^-}\quad , \quad
e^{2\f}={m\over \l} -\l^2 x^+ x^-\ .
\eqn\ii$$
Here $m$ is the black hole mass, and the $m=0$ solution
is called the linear dilaton vacuum (LDV). Note that
$g=e^\f$ is the coupling constant.

The goal then is to quantize the theory described by this
action $\Scl$. The $N$ matter fields give rise to the
well-known conformal anomaly
\foot{A possible term
$\mu^2\int \rg$ is supposed to be fine-tuned to vanish.}
$$\Sanom = -{\k\over 8\pi} \ix \rg R{1\over \nabla^2} R
\eqn\iii$$
with $\k={N\over 12}$. Note that this is ${\cal
O}(e^{2\f})\equiv {\cal O}(g^2)$ with respect to the
gravitational part of $\Scl$ and may be thought of as the
one-loop contribution of the matter fields. We will refer to
$\Scl + \Sanom$ as $S_{\rm CGHS}$. Evidently, we also have to
quantize the dilaton-gravity sector. In order to do so it
is convenient to go to conformal gauge. It was argued in
\REF\BC{A. Bilal and C. Callan, ``Liouville models for black
hole evaporation", Princeton University preprint PUPT-1320
(May 1992), to appear in Nucl. Phys. B,
hep-th@xxx/9205089.}  \REF\ALW{S. de Alwis,
 \PL B289 1992 278 ,
 \PL B300 1993 330 .}
refs \BC, \ALW\ that quantum
consistency requires the resulting theory to be a
conformal field theory. It turned out that this (almost
uniquely) determined the quantum-corrected action.
Moreover, when written in terms of appropriately redefined
field variables, the quantum-corrected action is very
simple and the corresponding \eoms can be solved exactly.
The non-trivial physics comes from the (transcendental)
transformation back to the ``physical" dilaton and metric
fields. We will review these theories briefly in section 2.

These exactly-solvable quantum-corrected theories seemed
to have no lower bound to the total energy (mass).\foot{
The same objection of course applies to the classical
theory \i, but since the solutions \ii\ are static and
non-radiating there will be no transitions from positive
to negative mass. In the CGHS model including $\Sanom$
there will be, a priori, such transitions, but since the
theory is not exactly solvable we cannot study them to
the same extent.}
The static solutions of negative mass have naked
singularities, just as the four-dimensional Schwarzschild
geometry. This, by itself, is not really worrisome, as
long as we can avoid that an initially singularity-free
solution dynamically develops a naked singularity. The
cosmic censorship conjecture (in 4D general relativity)
states that this can indeed be avoided. In the present
context of the 2D quantum-improved dilaton-gravity
theories, Russo, Susskind and Thorlacius (RST)
\REF\RST{J. Russo, L. Susskind and L. Thorlacius,
\PR D46 1992 3444 .}
\REF\CC{J. Russo, L.
Susskind and L. Thorlacius,
 \PR D47 1993 533 .}
[\RST,\CC] have shown that one can impose boundary
conditions so as to avoid naked singularities. (This
also applies to ref. \BC.) This results in matching an
evaporating black hole onto a vacuum configuration precisely
when the singularity is about to develop.

What is the mass of the (dynamical) black hole just
before it is matched to the vacuum? In their initial work
[\RST] RST claimed that it was ``slightly" (i.e. of order
$\k\l$) negative, and that this negative amount of energy is
sent off by a ``thunderpop". In fact, there seem to be (at
least) two different definitions (differing by $\Or
(\k\l)$-terms) for the mass of the black hole \REF\LAR{L.
Thorlacius, private communication.} [\LAR] one of them
leading to a slightly positive and one to a slightly
negative mass   just before the black hole disappears.

More generally we would like to have a theorem that (at
least one reasonable definition of) the mass or total energy
is positive as long as no singularity is encountered.\foot{
By singularity we do not simply mean a curvature
singularity, but a region of ``space-time" where the
dilaton gets complex. Of course, in general the curvature
diverges at the boundary of such a region. Strictly speaking
such regions should not be considered as part of the physical
space-time. In the four-dimensional analogue these
regions correspond to negative radius.}
Such a theorem was
proven by Park and Stominger
\REF\PS{Y. Park and A.
Strominger, ``Positive mass and supersymmetry in classical
and quantum two-dimensional dilaton-gravity", Santa Barbara
preprint UCSBTH-92-39, hep-th@xxx/9210017.}
[\PS] for the classical theory \i\
and for the CGHS-theory. They showed how to extend $\Scl$
to a supersymmetric theory and derived a spinorial
expression $M$ (that coincides under certain assumptions
with the conventionally defined total energy) from the
supercharge. This $M$ then is shown to be positive using the
\eoms of the bosonic theory given by $\Scl$. Thus although
supersymmetry is probably the underlying reason that makes
things work out, the positive energy proof itself does not
require supersymmetry, only the bosonic \eoms. Park and
Strominger then extend this proof to the CGHS theory by
representing the anomaly action \iii\ in a local form using
the ``$Z$"-field.

Here we will extend their work to the exactly solvable
quantum-corrected theories and proove the same result (for
$\k>0$):  {\it A
suitably defined mass functional $M$ given by an integral
over a space-like or null surface $\Sigma$ is always
non-negative as long as the dilaton-field is real on
$\Sigma$.} Now, the scalar curvature diverges at the
boundary of a region of complex dilaton field and is
complex inside. In all physically interesting situations
$\Sigma$ includes at least a portion where the dilaton is
real (e.g. an asymptotically flat region). Thus we can
conclude that {\it $M$ is non-negative as long as there is no
curvature singularity on $\Sigma$.} We also prove a
uniqueness theorem: under the assumptions specified below
our positive mass/energy functional is unique.

We will show how our energy functional when evaluated with
a space-like surface $\S$ leads to the usual ADM-mass plus
a certain ``quantum"-correction. When defined over a
null-surface $\S$ of constant $\sm$ we obtain an expression
for the Bondi-mass $\mb(\sm)$. The latter is shown to
behave as physically expected. We evaluate $\mb$ in detail
for the case of an evaporating black hole formed from an
infalling shock-wave. In particular, we show that at
$\sm=\sms$ when the singularity is about to develop and the
configuration is matched to the LDV, its Bondi-mass is
``slightly" {\it positive} and non-vanishing.

The total energy functional $M$ is again given by
some spinorial expression but our proof will only rely on
the (bosonic) \eoms. The spinors are just some
book-keeping device determined in terms of the metric and
the dilaton. One might however ask whether the
quantum-corrected theories under consideration have a
supersymmetric extension or not. This is particularly
interesting in view of the negative statement of Nojiri
and Oda
\REF\NOJ{S. Nojiri and I. Oda, \MPL A8 1993 53 .} [\NOJ].
Following Park and Strominger [\PS] it is easy to
explicitly write down these extensions although the
supersymmetric extensions of the cosmological-constant
term are rather non-trivial. For reasons first discussed in
\REF\DAN{U. Danielsson, ``Super black holes", CERN preprint
CERN-TH.6817-93, hep-th@xxx/9303002.}
ref. \DAN\ and repeated below, it is not the same to
construct a (classically) supersymmetric extension of a
given exact conformal theory (what we did) and to construct
an exact superconformal theory (what ref. \NOJ\ attempted
to do). This explains the apparent discrepancy with ref.
\NOJ.

The outline of this paper is as follows: in the next
section, we briefly review the
exactly-solvable quantum-corrected dilaton-gravity theories
and show how they can be rewritten using the ``$Z$"-field.
The expert reader might choose to skip part or all of this
section. In section 3, we give the mass functional $M$ and
show that it is non-negative provided the dilaton is real on
$\Sigma$. This is done by relating it to an expression
involving (a part of) the ``matter" stress-tensor which
is manifestly non-negative. We further show that, under
certain reasonable assumptions, this mass functional is
uniquely determined. In particular there is no freedom to
change terms that are subleading in the coupling constant
$e^\f$. In section 4, we show in some detail how this mass
functional $M$ is related to the usual Bondi and ADM
masses, discuss some of its properties and evaluate it in
particular for the shock-wave scenario. In section 5, we
write down the supersymmetric extensions of the
quantum-corrected theories of refs. \BC\ and \RST.

{ \chapter{The exactly solvable quantum-corrected
dilaton-gravity theories}}

Adding the matter anomaly piece \iii\ to the classical
action \i\ was a first step towards quantizing
two-dimensional dilaton-gravity [\CGHS]. The resulting
theory has two drawbacks, however. On the one hand, the
action $S_{\rm CGHS}$ is conformally invariant (after
shifting $\k$) only classically, but not at the quantum
level. On the other hand, the \eoms are not solvable in
closed form, which makes it difficult to study the dynamical
evolution. It turned out that solving the first problem also
cured the second: making the action conformally invariant,
at the same time leads to exactly solvable equations of
motion. Let us outline how this works.

First of all, not only the matter fields contribute to
the conformal anomaly, and as a result $\k$ is shifted to
[\BC]
$$\k={N-24\over 12}\ .
\eqn\dii$$
More generally, we expect other ${\cal O}( (
e^{2\f})^0 )$ corrections to $\Scl$ (which
itself is ${\cal O}( (
e^{2\f})^{-1} )$). It was shown in ref. \BC\
that in conformal gauge and after a local field
redefinition the kinetic part of $\Scl +\Sanom$ takes on
a free-field form. Then it was easy to identify the
correction to the cosmological constant term ($\sim
\l^2$) necessary to turn it into a marginal
operator.
The resulting theory was shown [\BC] to be a conformal
field theory (at the quantum level). It is given by
\foot{
Our notation is as usual: $\s^\pm=\s^0\pm\s^1\equiv
\tau\pm\s$, $\d_\pm =\half (\d_0\pm\d_1)$ and conformal
gauge is defined by $g_{++}=g_{--}=0,\, g_{+-}=-\half
e^{2\r}$, hence $\nabla^2=-4e^{-2\r}\dpl\dm$ on any
scalar.}
(not
writing the matter part $S^{\rm M}={1\over 2\pi}\is \sum_i
\dpl f_i \dm f_i$ explicitly)
$$S^{\r,\f}={1\over \pi} \is \left[ \k \dpl \O\dm\O
-\k\dpl\x\dm\x +\l^2 e^{2(\x-\O)}\right]
\eqn\dii$$
where $\x$ and $\O$ are the new fields related to the
dilaton $\f$ and the conformal factor of the metric
(Liouville field) $\r$ by\foot{We have rescaled $\O,\x$ by a
factor of 2 and shifted $\O$ by a constant with respect
to ref. \BC.}
$$\eqalign{\O&=\o\sqrt{\o^2-1}-\log\left(
\o+\sqrt{\o^2-1}\right) +\half\left( 1-\log{\k\over
4}\right) \quad , \quad \o=e^{-\f}/\sqrt{\k}\ ,\cr
\x&=\r+\o^2\ .}
\eqn\diii$$
This is valid for $\k>0$. See ref. \BC\ for the
appropriate analytic continuation to $\k<0$.

Alternatively, as in ref. \RST, one may modify the
kinetic part of $\Scl+\Sanom$ by adding an
${\cal O}( (
e^{2\f})^0 )$ correction
$$\delta S_{\rm RST}=-{\k\over 4\pi}\ix \rg\, \f R\ .
\eqn\diiia$$
This modifies the stress-tensor in such a way that the
(original) cosmological constant term ${1\over 2\pi} \ix
\rg e^{-2\f} 4\l^2$ becomes marginal. The resulting
action (in conformal gauge) can again be written in terms
of new fields $\x$ and $\O$ so that the resulting action
is identical to \dii, but the relation with the original
$\f$ and $\r$-fields is different (we have rescaled $\O$
and $\x$ by $\sqrt{\k}$ with respect to ref. \RST )
$$\eqalign{\O&={e^{-2\f}\over \k}+{\f\over 2}\ ,\cr
\x&={e^{-2\f}\over \k}-{\f\over 2}+\r\ .}
\eqn\div$$
Also, in terms of $\O$ and $\x$ the stress-tensor looks
identical in both cases [\BC,\RST]:
$$T_{\pm\pm}^{\r,\f}=\k\left[\left(\d_\pm\O\right)^2
-\left(\d_\pm\x\right)^2+\d_\pm^2\x\right]\ .
\eqn\dv$$
All this should be no surprise: there are not that many
conformal field theories with a canonical kinetic term
one can write down.

Although the action \dii\ is almost trivial - as shown
first in ref. \BC\ the \eoms can be solved exactly - the
field transformations \diii\ and \div\ certainly are not.
For $\k>0$ they become singular for some critical value
of the dilaton field $\f=\f_{\rm cr}$ where $\d\O/\d\f =
0$ and $\O$ is minimum. (We have $e^{-2\f_{\rm cr}}=\k$
for \diii\ and $e^{-2\f_{\rm cr}}=\k/4 $ for \div.)
In general, this corresponds to a real singularity of
the geometry, i.e. the scalar curvature diverges.\foot{
The LDV is the only exception: the curvature vanishes
everywhere and there is no singularity although $\f=\f_{\rm
cr}$ somewhere.} The dilaton
$\f$ is complex beyond the line of singularity which might be
interpreted as the boundary of physical space-time.

In the next section we will give a positive energy proof
that does not rely on conformal gauge but is generally
covariant. Now, although the anomaly action \iii\ is
local in conformal gauge, it is non-local when written in
the covariant form \iii. We will need a reformulation of
these exactly solvable quantum-corrected theories that is
local and covariant at the same time. We now proceed to
give such a reformulation, using the example of the
RST-action $\Scl + \Sanom + \delta S_{\rm RST}$. Define
\REF\ST{L. Susskind and L. Thorlacius, \NP B382 1992 123 .}
[\ST]    %
$$S_Z={1\over 2\pi} \ix \rg\left[ -\half (\nabla Z)^2
+QRZ\right] \ .
\eqn\dxiv$$
The $Z$-stress-tensor obtained from this action is [\PS]
$$\eqalign{\tmn^Z&=\hat \tmn^Z +Q\left( \N_\m\N_\n
Z-g_{\m\n}\N^2Z\right)\ ,\cr
\hat \tmn^Z &={1\over 2} \N_\m Z \N_\n Z-{1\over 4}
g_{\m\n}(\N Z)^2\ .}
\eqn\dxv$$
If we write
$$Z=\zt - Q{1\over \N^2}R
\eqn\dxvi$$
we have
$$S_Z=-{1\over 4\pi}\ix\rg (\N \zt)^2 - {Q^2\over 4\pi}
\ix\rg R{1\over \N^2} R\ ,
\eqn\dxvii$$
which is a free-field action for $\zt$ plus $\Sanom$ of
eq. \iii\ provided we take
$$2Q^2=\k\ .
\eqn\xviii$$
Note that $Z$ will be real only if $\k>0$.
Thus we want to consider
$$S=\Scl + \delta S_{\rm RST} + S_Z\ .
\eqn\newi$$
Naively, one might think that integration over $Z$, i.e.
over $\zt$ will just reproduce the anomaly action
$\Sanom$ and dispose of $\zt$ leaving us with the complete
dilaton-gravity part of the RST-action (i.e. excluding the
matter part). This is of course not the case since the
$\zt$-field is coupled to the other fields via the
$g_{\m\n}$-\eoms, i.e. the $T_{\pm\pm}=0$ constraints after
adopting conformal gauge. This is much the same as for the
matter fields $f_i$. Indeed, Park and Strominger  [\PS]
suggest to identify $T^Z$ with $T^{\rm M}+T^\r_{\rm anom}$
which
amounts to identifying $T^\zt_{\pm\pm}$ with $T^{\rm
M}_{\pm\pm}+t_\pm$. ($t_\pm$ is a projective connection
contained in $T^\r_{\rm anom}$ describing the boundary
values of the ``anomalous" stress-energy flux.)
With this identification it is easy to see that the dilaton
and $g_{\m\n}$-\eoms of the action $S$, eq.  \newi, are
rigorously identical to those of the original RST-action
(including the matter part).\foot{
Note that from $\tmn^Z$ we see that the $\zt$-field
contributes a (classical) conformal anomaly of $c_\zt =24
Q^2=12\k=N-24$.
Hence the $\zt$-field mimics not only
the matter fields but also the ghosts
plus certain quantum parts of $\f$ and $\r$. However, this
interpretation of the {\it fields} themselves should not be
taken too literally.}
Thus the RST action is equivalent to
$$S={1\over 2\pi}\ix \rg\left[ \left( e^{-2\f}-{\k\over
2}\f\right) R+e^{-2\f}\left( 4(\nabla\f)^2+4\l^2\right)
-\half (\nabla Z)^2 +QRZ\right]\ .
\eqn\dxxii$$
Going to conformal gauge and splitting
$$Z=\zt + 2Q\r\ ,
\eqn\dxxiii$$
we have
$$S={1\over \pi} \is \left[ \k \dpl \O\dm\O
-\k\dpl\x\dm\x +\l^2 e^{2(\x-\O)}+\half
\dpl\zt\dm\zt \right] \ ,
\eqn\dxxiv$$
where $\O,\, \x$ are given by \div, while the constraints
are
$$T_{\pm\pm}=0\quad ,\quad
T_{\pm\pm}=T_{\pm\pm}^{\r,\f}+T_{\pm\pm}^\zt\ ,
\eqn\dxxv$$
where $T_{\pm\pm}^{\r,\f}$ is still given by \dv\ and
$$T_{\pm\pm}^\zt =\half (\d_\pm \zt)^2+Q\d_\pm^2\zt\ .
\eqn\dxxvi$$
Let us repeat again that, upon identifying
$T_{\pm\pm}^\zt$ with $T_{\pm\pm}^{\rm M} +t_\pm$, eqs.
\dxxiv-\dxxvi\ are exactly equivalent with the RST model
(or that of ref. \BC).

For later reference, let us review one solution of the
\eoms in more detail. Since they are exactly solvable it
was  possible [\BC,\RST] to study the formation of a
black hole and its subsequent evaporation exactly,
automatically including the correct backreaction of the
Hawking radiation on the geometry. It turned out (as first
discussed in ref. \RST\ but also applying to ref. \BC) that
after the black hole formed the singularity is space-like
and hidden behind an apparent horizon. Then, as the black
hole evaporates, the apparent horizon recedes, until it hits
the singularity which then turns time-like and naked. At
this point one has to impose boundary conditions and it was
shown [\RST,\CC, Note Added to \BC] that one can match the
solution to a linear dilaton vacuum (static in some new
coordinates) or its analogue. If one does so, beyond that
point there is no more Hawking radiation and the black
hole has disappeared.

Since we will need it below, we will write out the field
configuration for this scenario. To be specific, we will
use the RST variant [\RST] since the transformation from
$\f$ and $\r$ to $\O$ and $\x$ is simplest here. The
asymptotically flat static solutions to the \eoms are
given by
$$\O(\f)={e^{2\l\s}\over \k}+2P\l\s +{m\over \l\k}\ .
\eqn\dvia$$
The LDV, $\f=-\l\s$, corresponds to $P=-{1\over 4},\, m=0$.
We use coordinates $\s^\pm=\t\pm\s$ such that
$\x=\O+\l\s$ or
$$\r=\f+\l\s\ .
\eqn\dviia$$
In these coordinates one has for the LDV solution
$\r=0$, i.e. the coordinates are
Minkowskian. In general these coordinates are
asymptotically Minkowskian coordinates as $\s\to\infty$..
In addition to the $\f,\r$-\eoms we must also satisfy the
$T_{\pm\pm}=0$ constraints (which are the
$g_{\pm\pm}$-\eoms before going to conformal gauge). They
are
$$T_{\pm\pm}=0\quad ,\quad T_{\pm\pm}=T_{\pm\pm}^{\r,\f}
+T_{\pm\pm}^{\rm M}+t_\pm
\eqn\dviiia$$
where $T_{\pm\pm}^{\rm M}$ is the matter stress-tensor
and $t_\pm(\s^\pm)$ is required to vanish in
asymptotically Minkowskian coordinates: $t_\pm(\s^\pm)=0$.
This and the constraints imply $P=-{1\over 4}$ for
$T_{\pm\pm}^{\rm M}=0$. $P\ne -{1\over 4}$ is appropriate
only if the solution is in equilibrium with a bath of
radiation.

Now let's suppose that we have a matter shock-wave\foot{
This is the analogue of a collapsing shell of matter in
four dimensions.} travelling along the line $\sp=\spo$
with stress-tensor
$$T_{++}^{\rm M}(\sp)=m \delta(\sp-\spo)\quad ,
\quad T_{--}^{\rm M}(\sm)=0\ .
\eqn\dixa$$
We use $\l x^\pm =\pm e^{\pm\l \s^\pm}$ and
$m=a e^{\l\spo}$. Equation \dixa\ corresponds to
$T_{++}^{\rm M}(x^+)=a\delta(x^+-x^+_0)$.
Note that by our preceeding discussion of the $\zt$-field
this $T_{++}^{\rm M}(\sp)$ corresponds to
$$
\dpl\zt=-2Q {\theta(\sp-\spo)\over \sp-\spo+{\k\over m}}\ .
\eqn\dixb$$
 We further take
the LDV solution for $\sp<\spo$, i.e. in the causal past
of the shock-wave trajectory. Matching across the
$\sp=\spo$ line (and requiring LDV asymptotics at right
past null infinity) leads to
$${e^{-2\f}\over \k} + {\f\over 2}\equiv \O(\f) = {1\over
\k} e^{\l\sp-\l\sm}-{\l\over 4}\sp+{\l\over 4}\sm
-{a\over \l\k} \left( e^{\l\sp}-\l x_0^+\right)
\theta(\sp-\spo)\ .
\eqn\dxa$$
The curvature is singular on the line where $\O=\O_{\rm
cr}$ (except when $\f$ is the LDV), where
$\O_{\rm cr}={1\over 4}\left(1-\log{\k\over 4}\right)$ is
the value of $\O(\f)$ at its minimum. As discussed in
ref. \RST, just above the infall-line $\sp=\spo$ the line
of singularity is space-like and hidden to the
asymptotically flat region $\s\to\infty$ by an
apparent horizon. As time goes on ($\t$ increases) the
apparent horizon and the line of singularity approach
each other until they intersect at $\s^\pm = \s^\pm_s$
where
$$\eqalign{
e^{\l\s^+_s}&={\k\l\over 4a}\left( e^{{4m\over \l\k}}-1
\right)\ ,\cr e^{\l\s^-_s}&={\l\over a}\left( 1-
e^{-{4m\over \l\k}}\right)\ .}
\eqn\dxia$$
The singularity turns time-like and naked. As shown by
RST, this can be avoided by matching the solution \dxa\ to
a shifted LDV for $\s^\pm > \s^\pm_s$. Indeed, on the
half-line $\sm=\sm_s,\ \sp>\sp_s$ the solution $\f$ given
by \dxa\ takes on LDV values:
$$\f=-{\l\over 2}\sp+{\l\over 2}\smt_s\ ,
\eqn\dxiia$$
where
$$\smt=\sm-{1\over \l}\log\left( 1-{a\over \l}
e^{\l\sm}\right)\ .
\eqn\dxiii$$
One chooses to take $\f={\l\over 2}\sp-{\l\over 2}\smt$
in all of the causal future of $(\sp_s,\sm_s)$.\foot{
The line where $\O=\O_{\rm cr}$ continues through this
shifted LDV (just as it was also present in the LDV
region $\sp<\spo$), but as in any LDV, this line does not
correspond to singular curvature, but rather to $R=0$.
One might however choose to consider as physical space-time
only the region to the right of this line.}

{ \chapter{The positive energy theorem}}

In this section, we will define a functional $M$ given by
an integral over a space-like or null surface $\S$ of a
suitable expression involving the fields. We will then
prove, for $\k>0$ (which we assume throughout this paper),
using the \eoms, that this functional $M$ is non-negative if
$\S$ is contained in the physical space-time, i.e. if $\f$
and $\r$ are real everywhere on $\S$. The expression for $M$
and the proof are suggested by Park and Strominger's analysis
[\PS] of the simpler CGHS-case.
In the next section, we will actually evaluate this
functional $M$ and show that it gives a satisfactory
Bondi-mass (for $\S$ a null line) and that (for space-like
$\S$) it reproduces the usual expression for the ADM-mass
plus a certain ``quantum"-correction.

We will write down the mass functional $M$ in covariant
form and use the covariant \eoms to prove positivity.
This is why we had to bother about rewriting the
non-local covariant anomaly term in a local form by
introducing the $Z$-field. All we will need are the
$g_{\m\n}$-\eoms of the action \dxxii. (We discuss the
RST variant here, since the algebra is simpler.) They are
$$T_{\m\n}^{g,\f}+T_{\m\n}^Z=0\ ,
\eqn\ti$$
where $T_{\m\n}^Z$ is given by \dxv\ and
$T_{\m\n}^{g,\f}$ is the covariant form of
$T_{\pm\pm}^{\r,\f}-T_{\pm\pm}^{\r,{\rm anom}}$, namely
$$\eqalign{T_{\m\n}^{g,\f}=&-2\left( e^{-2\f}+{\k\over
4}\right) \left( \N_\m\N_\n\f-g_{\m\n}\N^2\f\right)\cr
&-2 e^{-2\f}g_{\m\n}\left( (\N\f)^2-\l^2\right)\ .}
\eqn\tii$$

Given the structure of the \eoms one can easily guess how
the mass functional of ref. \PS\ should be modified in the
present case. Let
$$M=\int_\S {\rm d}\s^\m \, \N_\m\left[
2\left( e^{-2\f}+{\k\over 4}\right)
\eb\gf(\NS\f-\l)\e-Q\eb\gf\NS Z\e\right]\ ,
\eqn\tiii$$
where $2Q^2=\k$. $\e$ is a commuting real two-dimensional
spinor, and $\NS=\g^\m\N_\m=e^{\m a}\G_a\N_\m$, $e^{\m
a}$ being the zwei-bein and $\G_a$ Minkowski-space
Dirac-matrices obeying $\{\G_a,\G_b\}=2\eta_{ab}$. A
convenient choice that we adopt here is $\G_0=i\s_y,\,
\G_1=\s_x$. Let $\G_5=\G_0\G_1=\s_z$ while (following ref.
\PS) $\gf=\g^0\g^1=-\G_5$. As usual, $\eb=\e^+\G_0$. We will
also use the antisymmetric tensor normalized as
$\e_{0\phantom{1}}^{\phantom{0}1}
=\e_{1\phantom{0}}^{\phantom{1}0}=-1$.

The functional $M$ is given by a line integral of a
derivative along this line and thus reduces to the
difference of the values of the expression in the square
brackets at ``both ends of the world". Thus $M$ is given by
the asymptotic values of the fields and of the spinor
$\e$.\foot{
As discussed below, $\e$ is a solution of a
differential equation and, in general, its asymptotic value
depends on the values of the fields on all of $\S$. This
differs from 4D general relativity.}
In the next section, we will  discuss  under which conditions
this reproduces the more standard definition of mass in
terms of asymptotic field variations with respect to some
reference (vacuum) configuration.

We will prove that the above defined functional $M$ equals
$$M=\int_\S {\rm d}\s^\m \left( {1-{\k\over 4}e^{2\f}\over
1+{\k\over 4}e^{2\f}}\right)^2
\e_{\m\phantom{1}}^{\phantom{0}\r}\, \hat T^Z_{\r\n} \,
\eb\g^\n\e
\eqn\tiv$$
provided the spinor $\e$ satisfies the (ordinary)
differential equation
$${\rm d}\s^\m \left[\left( 1+{\k\over 4}e^{2\f}\right)
\N_\m \e -\half \g_\m (\NS\f-\l)\e -{Q\over 4}
e^{2\f}\left( 1+{\k\over 4}e^{2\f}\right)^{-1} \g_\m \NS Z
\e\, \right] =0\ .
\eqn\tv$$
This determines $\e$ only up to two functions of
integration. They are not relevant to the positivity
proof, but have to be specified to obtain a
physical interpretation of $M$ as the mass. This will be
done in the next section. The form \tiv\ of $M$ is manifestly
non-negative for $\k>0$ if $\f$ is real everywhere on $\S$.
Indeed, it is easy to see that for any real non-zero
$\e=\pmatrix{\e_1\cr\e_2\cr} $ (not necessarily a solution
of \tv), $v^a=\eb\G^a\e$ is time-like or null:
$v^av_a=-4(\e_1\e_2)^2\le 0$, and future-directed:
$v^0=\e_1^2+\e_2^2 >0$. The same then is obviously true for
$v^\m$. Now $\hat T^Z_{\m\n}$ (cf. \dxv) obeys the dominant
energy condition, i.e. for time-like or null,
future-directed $v^\n$ the vector $-\hat T^Z_{\m\n}v^\n$ is
again time-like or null, future-directed. Note that this is
true only if $Z$ is real, i.e. for $\k>0$! Indeed, for real
$\N_a Z=(f,g)$ we have $\hat T^Z_{00}={1\over
4}(f^2+g^2)=\hat T^Z_{11}\ge 0,  \hat T^Z_{01}=\half f g$
and obviously then $(\hat T^Z)^a_b v^b$ is a time-like or
null, future-directed vector. Since
$\e_{1\phantom{1}}^{\phantom{0}0}=-1$ it follows that $M$ as
given by \tiv\ is non-negative provided $\S$ is space-like
or null and $\f$ real on $\S$.

Let us now proceed to prove the equality of \tiii\ and
\tiv\ using \ti\ and \tv. We start with expression \tiii\
and evaluate $\N_\m [\ldots]$. Note that since we work
with real spinors,
$(\N_\m\eb)\gf\NS\f\e=\eb\gf\NS\f\N_\m\e$, etc. One uses
eq. \tv\ to get rid of all derivatives of $\e$. Employing
further the identities $\eb\gf\g^\m\e=
\e^{\m\phantom{1}}_{\phantom{0}\n}\eb\g^\n\e$ and $
\e^{\r\phantom{1}}_{\phantom{0}\n} a_\m a_\r=
\e_{\m\phantom{1}}^{\phantom{0}\r} (g_{\n\r} a^2
-a_\n a_\r)$
one arrives at
$$\eqalign{M=\int_\S {\rm d}\s^\m\  \eb\g^\n\e\,
\e_{\m\phantom{1}}^{\phantom{0}\r} &\Bigg[
2\left( e^{-2\f}+{\k\over 4}\right) (\N_\n\N_\r\f
-g_{\n\r}\N^2\f )\cr
&+2e^{-2\f}\, g_{\n\r}\left( (\N\f)^2-\l^2\right)\cr
&-Q\left( \N_\n\N_\r Z-g_{\n\r}\N^2 Z\right)\cr
&-{\k e^{2\f}\over \left( 1+{\k\over 4}e^{2\f}\right)^2}
\left( {1\over 2} \N_\n Z\N_\r Z-{1\over 4}g_{\n\r}(\N Z)^2
\right) \Bigg]\ .}
\eqn\tvi$$
By \tii\ and \dxv\ this equals
$$M=\int_\S {\rm d}\s^\m\,  \eb\g^\n\e\,
\e_{\m\phantom{1}}^{\phantom{0}\r} \Bigg[
-T^{g,\f}_{\n\r}-T^Z_{\n\r}+\hat T^Z_{\n\r}
-{\k e^{2\f}\over \left( 1+{\k\over 4}e^{2\f}\right)^2}
\hat T^Z_{\n\r}\Bigg]\ .
\eqn\tvii$$
Finally, by the \eom \ti\ this reduces to \tiv.

One might wonder whether the functional \tiii\ is the
only one for which one can prove positivity or whether
there are many others. We will show that \tiii\ is indeed
the only one {\it of this form}. More precisely, suppose
we start with a more general functional
$$\tilde M=\int_\S {\rm d}\s^\m  \,
\N_\m\left[\, \eb\gf\left( f_1(\f)\NS\f+f_2(\f)\l+f_3(\f)\NS
Z\right) \e\, \right] \eqn\tviii$$
with $\e$ subject to some more general first-order
differential equation
$${\rm d}\s^\m\left[ \N_\m\e-\g_\m\left( g_1(\f)
\NS\f+g_2(\f)\l+g_3(\f)\NS Z\right) \e\, \right] = 0 \ .
\eqn\tix$$
Note that the square brackets in \tviii\ and \tix\ are the
most general (spinorial) expressions one can write down that
are covariant, local and have the correct dimension.
 To prove
positivity we must be able to reexpress $\tilde M$ as an
integral containing only $\hat T^Z_{\n\r}$ which is the only
(``matter") piece obeying the dominant-energy condition.
Using the \eoms \ti\ we can also write $\hat
T^Z_{\n\r}=-T^{g,\f}_{\n\r}-( T^Z_{\n\r}-\hat T^Z_{\n\r})$.
Thus we require that, using now only \tix\ and spinor
identities, one can express $\tilde M$ as
$$\tilde M
=\int_\S {\rm d}\s^\m\,  \eb\g^\n\e\,
\e_{\m\phantom{1}}^{\phantom{0}\r} \left[ F_1(\f) \hat
T^Z_{\n\r} +F_2(\f)\left( -T^{g,\f}_{\n\r}-( T^Z_{\n\r}-\hat
T^Z_{\n\r})\right)\right]\ .
\eqn\tx$$
Under these assumptions it is straightforward algebra to
show that the functions $f_i(\f), g_i(\f)$ and $F_i(\f)$
are uniquely determined to be as in \tiii, \tv\ and
\tvii, \tiv.

Let us sketch the proof. What we will precisely show is
that equating \tviii\ and \tx\ using only \tix\
fixes the functions $f_i, g_i$ and $F_i$ to be
$$\eqalign{
f_1=\pm f_2=2c(e^{-2\f}+\kk ) & , \ \ f_3=-cQ \cr
g_1=\pm g_2={1\over 2}(1+ \kk e^{2\f} )^{-1} & , \ \
g_3={Q\over 4}e^{2\f}(1+ \kk e^{2\f} )^{-2}\cr
F_1=c\k e^{2\f}(1+ \kk e^{2\f} )^{-2} & , \ \ F_2=c
}
\eqn\ai$$
where $c$ is a constant. On the one hand, we start with
\tviii\ and evaluate $\N_\m$ of the square bracket, and
eliminate all $\N_\m \e$ and $\N_\m \eb$ using \tix.
On the other hand, we substitute the explicit expressions
\tii\ and \dxv\ for the energy momentum tensors into \tx.
All we have to do then is to compare the coefficients of
independent terms, $\eb\gf\NS\f\g_\m\e,\ \eb\gf\N_\m\f\NS
Z\e$, etc. Some care has to be exercised since independent
looking terms may be related by spinor identities. We end up
with the following system of equations:
$$\eqalign{
2f_1g_2+2f_2g_1+f_2'=0\quad &, \quad f_3g_2+f_2g_3=0\cr
f_1g_3+f_3g_1=0\quad &, \quad f_3'=0 \cr
QF_2+f_3=0 \quad &, \quad F_1+8f_3g_3=0\cr
F_2-e^{2\f}f_2g_2=0 \quad &, \quad
2(e^{-2\f}+\kk)F_2-f_1=0 \cr
2e^{-2\f}F_2+2f_1g_1+f_1'=0 \quad &, \quad 4f_1g_1+f_1'=0 \
.\cr }
\eqn\aii$$
These are ten equations for only eight functions. Thus it
appears to be non-trivial that one can solve this system,
i.e. that one can prove a positive energy theorem at all.
However, we actually can solve \aii\ and the only
solutions are those given in \ai. The constant $c$ only
determines the overall normalization of the energy and is
irrelevant to the positivity proof. We must choose $c>0$
(otherwise replace $M$ by $-M$) and hence it can always be
absorbed into the normalization of the spinor $\e$ (which
has to be fixed anyhow). We can thus choose %
$$
c=1\ .
\eqn\aiii$$
We are left with the sign ambiguity of $f_2$ and $g_2$
which reflects the symmetry $\l\to -\l$. It is thus
irrelevant, too. We conclude that the functions $f_i, g_i$
and $F_i$ are necessarily as in \tiii, \tv\ and \tvii,
\tiv.

Although  the mass
{\it functional}  \tviii\ is uniquely determined, its actual
value depends on the boundary or initial conditions imposed
on the spinor $\e$ upon solving its differential equation
\tix. They will be fixed in the next section by imposing
physical requirements.

{\chapter{Physical interpretation and applications}}

Now that we have a (to a certain extent) unique
functional $M$ that is non-negative we would like to see
whether it defines a reasonable mass and compute it for
various physically interesting scenarios.
In particular, we will evaluate $M$ as defined in \tiii\ for
the case where the field configuration is asymptotic to the
LDV at both ends of $\S$. If $\S$ is a space-like line one
should obtain the ADM-mass while a null-line $\S$ should
lead to the Bondi-mass. We will discuss the latter in
considerable detail and apply it to the shock-wace scenario
where we show how our functional $M$ produces the
physically expected behaviour of the Bondi-mass.

\section{ADM-mass}

We will first compute $M$
for a space-like line $\S$ of constant $\t$. Denote the
expression in square brackets in \tiii\ by $\M$, i.e.
$$
\M(\t,\s)=
2\left( e^{-2\f}+{\k\over 4}\right)
\eb\gf(\NS\f-\l)\e-Q\eb\gf\NS Z\e
\eqn\taaa$$
so that
$$M(\t)=\M(\t,\s=\infty) - \M(\t,\s=-\infty)\ .
\eqn\txi$$
Choose conformal coordinates and assume LDV asymptotics,
i.e. let asymptotically as $\s\to \pm\infty$
$$\eqalign{
\f&\sim -\l\s+\delta\f\cr \r&\sim \delta\r\cr}
\eqn\txii$$
where $ \delta\f$ and $\delta\r$ vanish as $\s\to
\pm\infty$.  Let furthermore $\zt\to 0$
as $\s\to \pm\infty$ so that $Z\sim 2Q\r$.

The key point in computing the total energy $M$ is to
 specify  the asymptotics of the spinor $\e$ (and hence
also its normalization) . Since $\e$ is a solution of the
differential equation \tv\ it is determined by two
functions of integration which we may take to be determined
by the asymptotics of $\e$ at {\it one} end of $\S$ (e.g.
$\s\to +\infty$). Obviously then, to fix the two
functions of integration we need to specify the leading {\it
and} the {\it subleading} term in the asymptotic expansion of
$\e$ as $\s\to\infty$.

The LDV asymptotics as $\s\to +\infty$ implies
$$
\r\sim a_1(\t)e^{-\l\s}+a_2(\t)e^{-2\l\s}+\ldots\ ,\ \
\f=-\l\s+\r
\eqn\qqiia$$
where we use a suitable set of coordinates so that
$\f=-\l\s+\r$. It is related to the ``Kruskal" coordinates
$x^\pm$ where $\f=\r$ by the usual transformation
$\l x^\pm=e^{\pm\l\s^\pm}$. The \eoms and the constraints
together with the asymptotics \qqiia\ imply the following
relations:
$$
\dot a_2=2a_1 \dot a_1\ \ , \quad \ddot a_1 =\l^2 a_1\ .
\eqn\qqiiia$$
Asymptotically
the differential equation for $\e$ reads for a spacelike
surface $\S$ of constant $\t$:
$$
\d_\s \e=-{\l\over 2}(1+a_1 e^{-\l\s})(1+\G_1)\e + \Or (
e^{-2\l\s})\ .
\eqn\qqiii$$
If we write the solution as
$$
\e=\e_{(0)}+\e_{(1)}e^{-\l\s} + \Or (e^{-2\l\s})
\eqn\qqiv$$
then the differential equation implies
$$
(1+\G_1)\e_{(0)}\ =\ (1-\G_1)\e_{(1)}\ =\ 0
\eqn\qqv$$
i.e.
$$
\e_{(0)}={c_0\over \sqrt{2}}\pmatrix{1\cr -1\cr}\quad ,
\quad  \e_{(1)}={c_1\over \sqrt{2}}\pmatrix{1\cr 1\cr}
\eqn\qqvi$$
where $c_0$ and $c_1$ may depend on $\t$ and are the two
functions of integration. Note that \qqvi\ ensures that
$\M(\t,\s=+\infty)$ and hence $M$ does not diverge for
configurations asymptotic to the LDV. Given $c_0$ and
$c_1$, the differential equation completely determines $\e$,
and in particular its asymptotics as $\s\to -\infty$. The
latter however not only depends on the asymptotics of the
fields but on their values on all of $\S$.

For the LDV, the exact solution is ($\so$ is some fixed
reference point)
$$
\e_{\rm LDV}(\s)= {1-\G_1\over 2} \e(\s_0)
+\left( e^{2\l\so}+\kk\over e^{2\l\s}+\kk \right)^{1/2}
{1+\G_1\over 2} \e(\s_0)
\eqn\qqvii$$
from which one may read off
$\e_{(0)}={1-\G_1\over 2} \e(\s_0)$ and
$\e_{(1)}=\left( e^{2\l\so}+\kk\right)^{1/2}
{1+\G_1\over 2} \e(\s_0)$. Using the exact result \qqvii\ it
is  straightforward to obtain
$$\M_{\rm LDV}(\t,\s)=-2\l \left( e^{2\l\so}+\kk\right)
\eb(\so)\gf (1+\G_1)\e(\so)
\eqn\qqviii$$
which is independent of $\s$, and hence by \txi\
$$M_{\rm LDV}=0
\eqn\qqix$$
independent of the choice of $\e(\so)$, i.e. of the
functions of integration $c_0$ and $c_1$. It is worthwile
noting that unless $\e_{(1)}=0$ the total energy for the
LDV receives contributions from both ends of $\S$ (which
cancel each other). One sees that one has to take carefully
into account $\M(\t,\s=-\infty)$ as well as the subleading
term ($\sim \Or (e^{-\l\s})$) in $\e$ when evaluating
$\M(\t,\s=+\infty)$.

For the general LDV-asymptotic configuration \qqiia\ it is
straightforward to obtain
$$
\M(\t,\s=+\infty)=2\l ({a_1^2\over 2} - a_2) c_0^2 +4\dot
a_1 c_0c_1 -4\l c_1^2 \ .
\eqn\qqx$$
For $\s\to -\infty$, only the leading asymptotics
contribute, and since these are LDV asymptotics we can read
off the result from \qqviii\ if we take $\s_0$ in the
region where the LDV asymptotics is valid, i.e.
$\s_0=-\infty$ :
$$
\M(\t,\s=-\infty) =-{\k\over 2} \l \eb\gf(1+\G_1)\e
\vert_{\s=-\infty}\ .
\eqn\qqxi$$
Note that, by equation \qqv\ the terms $\Or (e^{-2\l\s})$
in the asymptotic expansion of $\e$ do not contribute
to $\M(\t,\s=+\infty)$. Due
to the explicit factor of $\k$ in $-\M(\t,\s=-\infty)$ one
might want to interpret the latter as a quantum correction to
$\M(\t,\s=+\infty)$. Looking at the LDV example, eq.
\qqviii, however, shows that this is not possible in general
and might only be true for some particular choice of $c_0$
and $c_1$.

We will now consider such a choice and set
$c_0=1$. The other function of integration, $c_1$ is
fixed by requiring
$$
\lim_{\s\to\infty} e^{-\f} (\NS\f-\l)\e=0 \ .
\eqn\qqxiia$$
Indeed, this fixes the subleading term $\e_{(1)}$ in
the expansion \qqiv\ of $\e$, since the leading term $\sim
(1+\G_1)\e_{(0)}$ vanishes by the differential equation,
see \qqv.\foot{
Note that ref. \PS \ requires
$\lim_{\s\to\infty} (\NS\f-\l)\e=0$ to fix $\e$. However,
as we have just seen, this is an empty statement, since it
cannot fix $\e_{(1)}$, while $(1+\G_1)\e_{(0)}=0$ as a
consequence of the differential equation, anyhow.}
Equation \qqxiia\ determines $c_1$ as $c_1 = {\dot a_1\over
2\l} c_0$.
Then, choosing $c_0=1$ fixes the normalisation. The
latter can be obtained  by requiring
$$\
\eb\gf\e\vert_{\s=\infty} =1\ .
\eqn\qqxiib$$
Remark, that an alternative choice would be to replace
\qqxiia\ by the following condition at $\s=-\infty$:
$(\NS\f-\l)\e\vert_{\s=-\infty}=0$. Then
$\M(\t,\s=-\infty)=0$ and $M=\M(\t,\s=+\infty)$. $c_1$ then
has to be obtained by solving the $\e$-differential equation
for all $\s$. This type of approach is advocated in the
next subsection for computing the Bondi-mass, and it could
also be carried out for the present discussion of the
ADM-mass.

At present, however, we will simply remark that
if we choose to impose \qqxiia\ and \qqxiib, i.e. $c_0=1$
and $c_1= {\dot a_1\over
2\l} c_0$, we obtain
$$
\M(\t,\s=+\infty)=\l a_1^2-2\l a_2+{\dot a_1^2\over \l} \ .
\eqn\qqxiva$$
Note that using \qqiiia\ we get
$${{\rm d}\over {\rm d}\t} \M(\t,\s=+\infty)=0\ .
\eqn\qqxivb$$
Thus we find that at least the contribution from
$\s=+\infty$ is time-independent.\foot{Although we were not able to prove
it, the contribution from $\s=-\infty$ should also be time-independent
in order to produce a
satisfactory definition of  ADM-mass.} Let us compare
\qqxiva\ with other expressions for the ADM-mass given in
the literature. First, for $a_1=0$ (and only in this case),
expression \qqxiva\ equals
\foot{ More precicely, from \qqiiia\ we see that $a_1(\t)
e^{-\l\s}\sim a_+ e^{\l\sp}+a_-e^{-\l\sm}$, and hence there
is a conformal coordinate transformation that sets $a_1$ to
zero in the new coordinates. Of course, in the new
coordinates, where $\delta\r$ and $ \delta\f$ are $\Or
(e^{-2\l\s})$, we no longer have $\f=-\l\s +\r$. Repeating
the above computation with $\delta\r \ne \delta\f$ one
obtains the desired equation.}
$$
\M(\t,\s=+\infty)=\lim_{\s\to\infty} 2e^{2\l\s} (\d_\s
\delta\f+\l\delta\r)
\eqn\qqxii$$
which is the standard expression for the ADM-mass usually
used in the literature
\REF\WIT{E. Witten, \PR D44 1991 314 .}
[\WIT,\CGHS] (it is correct only if $a_1=0$, since otherwise
it is {\it not} $\t$-independent). For general $a_1$,
expression \qqxiva\ equals $2\l (a_1^2-a_2) +\Delta$, where
$\l\Delta=\dot a_1^2-\l^2a_1^2$ is a constant by equation
\qqiiia. Up to the constant $\Delta$-term our expression
\qqxiva\ coincides
with the conserved expression for the mass,
$\lim_{\s\to\infty} 2e^{2\l\s}
(\d_\s+\l)(\delta\f-\delta\f^2)$,
 derived in
\REF\BK{A. Bilal and I.I. Kogan, ``Hamiltonian approach to 2D
dilaton-gravities and invariant ADM-mass", Princeton
University preprint PUPT-1379, to appear in Phys. Rev. D,
hep-th@xxx/9301119.} ref. \BK.

 With
these remarks in mind, we have shown that for a space-like
surface of constant $\t$ and for the more restricted
asymptotics ($a_1=0$), the energy functional $M$, with $\e$
subject to \qqxiia, \qqxiib, equals the usual ``classical"
ADM-mass plus a ``quantum"-correction $-\M(\t,\s=-\infty)$.
The latter by itself is  non-negative (since
$\eb\gf(1+\G_1)\e = \vert \e_1+\e_2 \vert^2 \ge 0$).
Obviously our positivity theorem  allows the classical
expression for the ADM-mass to get slightly negative by just
the amount that is compensated for by the
``quantum"-correction.

\section{Bondi-mass}

We will now consider a light-like surface $\S$ of constant
$\sm$ and obtain an expression for the Bondi-mass. Again,
we work in coordinates where $\f=-\l\s+\r$. First we solve
the $\e$-differential equation {\it exactly}. It reads
(with $\e=\pmatrix{\e_1\cr \e_2\cr}$)
$$\eqalign{
\dpl \e_1&=\left[ {\dpl\f\over 1+x}+\left( -{1\over 2} +
{2x\over (1+x)^2}\right)\dpl\r +{2Q\over \k} {x\over
(1+x)^2} \dpl \zt\right] \e_1 -{\l\over 2} {e^\r\over
1+x}\ \e_2\cr
\dpl \e_2&={1\over 2} e^\r  \e_2 \cr}
\eqn\qqxiii$$
where $x=\kk e^{2\f}$. Making use of $\r=\f+\l\s$ this is
readily integrated:
$$
\e\ =\ \pmatrix{
e^H\left[ d_1 - d_2 {\l\over 2}\int_{\spo}^{\sp} {\rm d}\spt
\left( 1+\kk e^{2\f}\right)^{-1} e^{{3\over 2}\r-H}\right]
\cr
e^{{1\over 2}\r} d_2 \cr
}\ .
\eqn\qqxiv$$
Here $\spo$ is some arbitrary reference coordinate, and the
function $H(\sp,\sm)$ is given by
$$\eqalign{
H=&-{\l\over 2}(\sp-\spo)+{1\over 2}\left[ \r-\log (1+x)
+{2x\over 1+x}\right]^{\sp}_{\spo} \cr
&+\int_{\spo}^{\sp} {\rm d} \spt {x\over (1+x)^2}\left( \l
+{2Q\over \k}\dpl \zt \right) \ .\cr}
\eqn\qqxv$$
$H$ vanishes at $\sp=\spo$ so that
$$
\e(\spo,\sm)\ =\ \pmatrix{ d_1(\sm)\cr
e^{{1\over 2}\r(\spo,\sm)} d_2(\sm) \cr }
\eqn\qqxvi$$
i.e. $d_1(\sm)$ and $d_2(\sm)$ are ``constants" of
integration.

Next, we need to study the asymptotics as $\sp\to
\pm\infty$. Consider $\sp\to +\infty$ first:
$$\eqalign{
\r&\sim a_0(\sm)+a_1(\sm)e^{-\l\sp}+\Or(e^{-2\l\sp})\cr
\f&=-{\l\over 2}\sp+{\l\over 2}\sm +\r\cr
\dpl\zt&\to 0\qquad\qquad \qquad\qquad  \qquad\qquad {\rm as}\ \sp\to
+\infty\ .}
\eqn\qqxvii$$
Note that we do {\it not} have $\r\to 0$ but rather $\r\to
a_0(\sm)$.
In order to be consistent with the LDV asymptotics as
$\s\to\infty$ one must have $a_0(\sm)\to 0$ as $\sm\to
-\infty$. This is satisfied for the shock-wave scenario
where $a_0=-{1\over 2} \log\left( 1-{a\over \l}
e^{\l\sm}\right)$ and also for more general solutions.
Indeed, the general solution of the \eoms and constraints
with a matter stress-energy $T^{\rm M}_{++}$ vanishing for
large enough $\sp$ (or decreasing sufficiently fast)
and $T^{\rm M}_{--}=0$ yields
$$\eqalign{
a_0&=-{1\over 2} \log\left( 1-{p\over \l}e^{\l\sm}\right)\cr
a_1&=-{1\over 2}{e^{\l\sm}\over 1-{p\over \l}e^{\l\sm}}
\left[ {m\over \l} +\kk
\log\left( 1-{p\over \l}e^{\l\sm}\right)\right] \ . }
\eqn\qqxviia$$
Here $m$ and $p$ are the total energy and momentum carried
by the infalling matter. (The shock-wave corresponds to
$p=a$ and $m=ae^{\l\spo}=a\l x_0^+$ in the usual notation.)
Note that $a_0$ satisfies the relation
$$
{2\over \l} a_0' +1-e^{2a_0}=0
\eqn\qqxviii$$
which we shall use below.
The asymptotics \qqxvii\ imply for $\sp\to\infty$
$$\eqalign{
H&=-{\l\over 2}(\sp-\spo)+H_0(\sm) +\Or (e^{-\l\sp})\ ,\cr
H_0(\sm)&={a_0\over 2}
-{1\over 2}\left[ \r-\log (1+x)
+{2x\over 1+x}\right]_{\spo} +\int_{\spo}^\infty {\rm d}
\spt {x\over (1+x)^2}\left( \l +{2Q\over \k}\dpl \zt \right)
} \eqn\qqixx$$
(recall that $x=\kk e^{2\f}$). Then the integrand
in eq. \qqxiv\ for $\e$
behaves as $(1+x)^{-1} e^{{3\over 2}\r-H} \sim \exp \left(
{3\over 2}a_0-H_0+ {\l\over 2}(\spt-\spo)\right) +
\Or (e^{-{\l\over 2}\sp})$. As a result, as $\sp\to\infty$:
$$\eqalign{
\e&=\e_{(0)}+\e_{(1)} e^{-{\l\over 2}\sp} +\e_{(2)}
e^{-\l\sp}+\ldots ,\cr
\e_{(0)}&=-e^{a_0} d_2 \pmatrix{ e^{{1\over 2}a_0}\cr
-e^{-{1\over 2}a_0}\cr }\ \ ,\quad
\e_{(1)}={1\over \sqrt{2}}e^{{3\over 2}a_0} L
\pmatrix{1\cr 0\cr} }
\eqn\qqxx$$
where
$$\eqalign{
L(\sm)&=\sqrt{2} e^{-{3\over 2}a_0+H_0 +{\l\over 2}
\spo}
\left[ d_1+d_2\left( e^{{3\over 2}a_0-H_0}-J\right) \right]
\ ,\cr
J(\sm)&={\l\over 2}\int_{\spo}^\infty {\rm d} \spt
\left[ \left( 1+\kk e^{2\f}\right)^{-1} e^{{3\over 2}\r-H}
-e^{{3\over 2}a_0-H_0+{\l\over 2}(\sp-\spo)} \right]\ .}
\eqn\qqxxi$$
At this point the reader might wonder why it looks so
complicated to extract the asymptotics of $\e$. Of course,
the {\it form} \qqxx\ of the asymptotics follows immediately
from the $\e$-differential equation. However, this does not
determine the value of $L$. We may take $L$ as a free
parameter, but then we need to use the exact
solution of the differential equation to obtain $\e$ at the
other end of $\S$ (as $\sp\to -\infty$). Our approach here
is to fix $\e$ at some finite $\spo$ (cf \qqxvi) and to
determine the value of $L(\sm)$. This will be particularly
convenient for any configuration that equals the LDV for
all $\sp < \spo$, as is the case e.g. in the shock-wave
scenario.

It is now straightforward to compute
$$
\M(\sm,\sp=+\infty)=\lim_{\sp\to\infty}
\left[ 2\left( e^{-2\f}+\kk\right)
\eb\gf(\NS\f-\l)\e-\k\eb\gf\NS \r\e\right]
\eqn\qqxxii$$
where we already used $\d_\pm\zt\to 0$ as $\sp\to\infty$.
Inserting the asymptotics \qqxvii\ and \qqxx, and using the
relation \qqxviii, one arrives at
$$
\M(\sm,\sp=+\infty)=-4d_2^2 e^{-\l\sm}e^{-2a_0} a_1'
-\k\l d_2^2 (1-e^{2a_0}) -\l e^{-\l\sm} L^2\ .
\eqn\qqxxiii$$
Using now the solution \qqxviia\ of the \eoms and
constraints one obtains
$$\eqalign{
\M(\sm,\sp=+\infty)=&
{2d_2^2(\sm)\over 1-{p\over \l}e^{\l\sm}}
\left[ m +\kk \l
\log\left( 1-{p\over \l}e^{\l\sm}\right)
+\kk p e^{\l\sm}\right] \cr
&-\l e^{-\l\sm} L^2(\sm)\ .\cr}
\eqn\qqxxiv$$

Next, we evaluate $\M(\sm,\sp=-\infty)$. We will do so
under the assumption that there is some value $\sp_*$ of
$\sp$ so that we have the LDV ($\r=0, \f=-{\l\over 2}\sp
+{\l\over 2}\sm, \dpl\zt=0$) for all $\sp<\sp_*$. We then
identify the (so far arbitrary) value of $\spo$ with this
$\sp_*$. In the LDV region ($\sp<\spo$) our formulas
simplify:
$$
H=-{1\over 2}\ \log\, {e^{\l(\sp-\sm)}+\kk \over
e^{\l(\spo-\sm)}+\kk }
\eqn\qqxxv$$
and then from equation \qqxiv\
$$
\e\ =\ \pmatrix{ -d_2\cr d_2\cr} +
\left( {e^{\l(\spo-\sm)}+\kk \over
e^{\l(\sp-\sm)}+\kk }\right)^{1/2} \pmatrix{ d_1+d_2\cr
0\cr} \ .
\eqn\qqxxvi$$
Note that this has a finite limit as $\sp\to -\infty$ for
all finite $d_1(\sm), d_2(\sm)$. We obtain
$$\eqalign{
\M(\sm,\sp=-\infty)&=\lim_{\sp\to -\infty} \left( - {\k\l
\over 2}\right) \eb\gf(1+\G_1)\e \cr
&=-2\l(d_1+d_2)^2 \left( e^{\l(\spo-\sm)}+\kk \right) }
\eqn\qqxxvii$$
and the Bondi-mass equals
$$\eqalign{
M_{\rm B}(\sm)&= \M(\sm,\sp=+\infty)-\M(\sm,\sp=-\infty)\cr
&={2d_2^2(\sm)\over 1-{p\over \l}e^{\l\sm}}
\left[ m +\kk \l
\log\left( 1-{p\over \l}e^{\l\sm}\right)
+\kk p e^{\l\sm}\right] \cr
&-\l e^{-\l\sm} L^2(\sm)
+2(d_1+d_2)^2 \l \left( e^{\l(\spo-\sm)}+\kk \right)\ . }
\eqn\qqxxviii$$

So far, $M_{\rm B}$ still depends on two arbitrary
functions $d_1(\sm)$ and $d_2(\sm)$. They have to be fixed
by physical requirements.
First of all, the LDV should have vanishing energy. It is
easy to check that this is true independent of the choice
of $d_1, d_2$:
$$
\mb^{\rm LDV}=0\ .
\eqn\qqxxix$$
Second, for $\k=0$, we do not expect any Hawking radiation
in the RST model and $\mb$ should equal $m$ for all $\sm$.
Since, for $\k=0$, we can solve for $\r$ and $\f$
explicitly, $\r=-{1\over 2} \log \left( 1-{p\over
\l}e^{\l\sm}+{m\over \l} e^{-\l\sp+\l\sm}\right),\
\f=-\l\s+\r$, we can evaluate $J$ and hence $L$ exactly.
The final result for $\mb$ is very simple:
$$
\mb\vert_{\k=0}=2d_1^2\vert_{\k=0}\, m\ .
\eqn\qqxxx$$
Thus we find $d_1^2={1\over 2}+\Or (\k)$.

Third, we
consider the general case corresponding to \qqxxviii\ in the
limit $\sm\to -\infty$. Again one should find
$\mb(\sm=-\infty)=m$ since no Hawking radiation yet had a
chance to be emitted. Evaluating carefully the asymptotics
(as $\sm\to -\infty$) of $\r, \f$, and using \dixb\ to
obtain those of $J$ and $L$, we
find
$$\eqalign{
\mb(\sm=-\infty)=& 2d_2^2(-\infty) +
(d_1(-\infty)+d_2(-\infty))^2 (2pe^{\l\spo}+2\k\l I_1)\cr
&-2 d_2(-\infty) (d_1(-\infty)+d_2(-\infty))
(2pe^{\l\spo}+\k\l I_{1/2})\cr }
\eqn\qqxxxi$$
where $I_1$ and $I_{1/2}$ are logarithm integrals defined
below. We will now argue that the only physically
acceptable choice is $d_1+d_2=0$, $d_2^2={1\over 2}$.
Indeed, if $d_1+d_2\ne 0$ the equation
$\mb(\sm=-\infty)=m$  defines $d_2(-\infty)$  as a
function of   $(d_1+d_2)(-\infty)$  in a way that
depends on $p$ and $m$. In particular, $\e$ would depend on
these quantities even in the LDV region which is in the
causal {\it past} of the infalling matter distribution.
Hence, we argue that by causality we should have
$d_1(-\infty)+d_2(-\infty)=0$, and hence
$d_2^2(-\infty)=d_1^2(-\infty)={1\over 2}$. This would still
leave the possibility that for finite $\sm$ we have
$d_1+d_2=\Or (e^{\l\sm})$. However, by the same argument,
in the LDV region, $\e$ should not depend on $\spo$ which
determines the trajectory of the shock-wave. Looking at the
exact solution \qqxxvi\ for $\e$ in the LDV region we see
that this implies $d_1+d_2=0$ for all $\sm$.
Thus we arrive at
$$
d_1+d_2=0\ ,\quad d_1^2=d_2^2={1\over 2}\ ,\quad \forall\
\sm\ ,
\eqn\qqxxxii$$
so that in the LDV region $\e=\pm {1\over \sqrt{2}}
\pmatrix{ 1\cr -1\cr}$. The conditions \qqxxxii\
 can be written more elegantly as
$$
e^{-2\r}\eb\gf\e\vert_{\sp=+\infty}=-1
\eqn\qqxxxiia$$
which fixes $d_2^2={1\over 2}$, and (note that the l.h.s.
is taken at $\sp=-\infty$, {\it not} $+\infty$)
$$
(\NS \f-\l)\e\vert_{\sp=-\infty}=0
\eqn\qqxxxiib$$
which fixes $d_1+d_2=0$.
With this choice, $\M(\sm,\sp=-\infty)=0$, and
$$
M_{\rm B}(\sm)
={1\over 1-{p\over \l}e^{\l\sm}}
\left[ m +\kk \l
\log\left( 1-{p\over \l}e^{\l\sm}\right)
+\kk p e^{\l\sm}\right]
-\l e^{-\l\sm} L^2(\sm)
\ .
\eqn\qqxxxiic$$

We now evaluate $\mb$ for large negative but finite $\sm$.
Keeping the first subleading terms as $\sm\to -\infty$, we
have
$$\eqalign{
J&\sim -{1\over 2}\left[ {p\over \l} e^{\l\spo} +\kk +\k
I_{1/2} -\k I_1 \right] e^{-\l\spo} e^{\l\sm}\ , \cr
L&\sim \pm \left[ {p\over \l} e^{\l\spo}
+{\k\over 2} I_{1/2} \right] e^{-{\l\over 2}\spo} e^{\l\sm}
}
\eqn\qqxxxiii$$
where $I_\alpha$ is a logarithm integral given by
$$
I_\alpha = e^{\alpha\l\spo}\int_{\spo}^\infty {\rm d} \sp
{e^{-\alpha\l\sp}\over \sp-\spo+{\k\over m} }
=\int_0^\infty {\rm d} x {e^{-x}\over x+{\alpha\l\k\over m}} =
-e^{{\alpha\l\k\over m}}\, {\rm li}\left(
e^{-{\alpha\l\k\over m}} \right)\ .
\eqn\qqxxxiv$$
Thus, for the shock-wave scenario ($pe^{\l\spo}=m$), we arrive
at
$$\mb(\sm) \sim m -\k\left( m I_{1/2}
+ {\l\k\over 4}
I_{1/2}^2 \right) e^{-\l\spo} e^{\l\sm} + \Or (e^{2\l\sm})\ .
\eqn\qqxxxv$$
Let us comment on this equation. First, as already
observed, $\mb$ is constant for $\k=0$: classically there
is no Hawking radiation. Second, $\mb$ is decreasing as
$\sm$  (i.e. time) increases (at least to the first order
in $e^{\l\sm}$ we computed): Hawking radiation carries
energy away from the black hole.\foot{ Recall that the
we assume $\k\ge 0$ throughout this paper.} Note that for
${\l\k\over m} << 1$ the leading term in \qqxxxv\ reads
$\mb \sim m -\k m\log\left( {2m\over \l\k}\right)
e^{-\l\spo} e^{\l\sm}$. This differs from the CGHS
prediction for the very early Hawking radiation by the
extra factor of $\log\left( {2m\over \l\k}\right)$.
However, there is nothing wrong with this difference, since
the RST and CGHS models represent different $\Or (\k)$
corrections to the same classical dilaton-gravity.

Finally we would like to compute $\mb(\sm)$ for the
shock-wave scenario at $\sm=\sm_s$, the point where the
singularity and apparent horizon intersect, and the
solution is matched to the LDV. As for any finite $\sm$, we
have no {\it explicit} functions for $\f$ and $\r$ (they are
given by solving the transcendental equation \dxa\ at each
point) which makes it difficult to obtain an exact
expression for $L$ since it involves integrals of
functions of $\f$ over all $\sp>\spo$. One could, of course,
proceed numerically. For $\sm=\sms$, however, the situation
is slightly better since we know that $\f$ and $\r$
equal the ``shifted" LDV for $\sp>\sps$ (cf. \dxiia):
$$
\f(\sp,\sms)=-{\l\over 2}(\sp-\sp_s)-{1\over 2} \log {\kk}\
,\quad  \r(\sp,\sms)={2m\over \l\k}\ ,\quad \sp>\sps\ .
\eqn\qqxxxvi$$
On the other hand, if ${m\over \l\k}<<1$, $\sps$ is close
to $\spo$ : $\l\sps-\l\spo={2m\over \l\k}+\Or( ({2m\over
\l\k})^2)$, and we have to solve for $\f$ and $\r$ on the
small interval $[\spo,\sps]$ only which can be done
perturbatively in ${m\over \l\k}$. It is easy to see that
all quantities can be developed in powers of ${m\over
\l\k}$ (e.g. no $\log {m\over \l\k}$ occurs contrary to the
opposite limit ${m\over \l\k}>>1$). Here we only compute
$\mb(\sms)$ to first order in ${m\over \l\k}$ which will
turn out to be very easy. Indeed, write
$L(\sms)=\alpha +\beta {m\over \l\k}+\ldots$. But we
know that in the limit where $m\to 0$ (i.e. in the LDV)
$L$ vanishes. Hence $\alpha=0$. Using the explicit expression
for $\sms$, equation \dxia, we find
$$\mb(\sms)={\k\l\over
4}\left[ \left( e^{{4m\over \l\k}}-1\right) - {{4m\over
\l\k}\over 1-e^{{4m\over \l\k}}} e^{-\l\spo}
L^2(\sms)\right]\ . \eqn\qqxxxviii$$
Expanding to first order in ${m\over \l\k}$, $L^2$ does not
contribute and
$$\mb\sim m +\Or( ({m\over \l\k})^2)\ .
\eqn\qqxxxix$$
Thus if we start with a very small black hole (small $m$)
or a very large number of matter fields (large $\k$), the
black hole is matched to the shifted LDV before any
substantial Hawking radiation has occurred: its mass is
still the initial mass $m$ up to $\Or( ({m\over \l\k})^2)$
corrections. This positive amount of energy must then be
sent off by the thunderpop. In ref. \RST, RST find (up to
their sign ambiguity) that the thunderpop carries energy
${\l\k\over 4}\left( 1-e^{-{4m\over \l\k}}\right)=m +
\Or( ({m\over \l\k})^2)$ in agreement with \qqxxxix.

In conclusion, we have found that our functional $M$ as
given by \tiii\ with $\e$ subject to the differential
equation \tv\ and the boundary conditions \qqxxxiia\ and
\qqxxxiib\ defines a satisfactory Bondi-mass: it is
non-negative, equals the ADM-mass $m$ at $\sm=-\infty$,
decreases for $\k>0$ and is constant for $\k=0$. It also
gives correctly the energy of the thunderpop (at least to
the order we computed).

{\chapter{ The supersymmetric extension}}

Positive energy theorems naturally occur in
supersymmetric theories. Thus the
preceeding results prompt the question: does there exist
a supersymmetric extension of the action \dxxii\ or \dxxiv
? For the CGHS model such an extension was constructed in
ref. \PS. There it was also shown that starting from a
general supersymmetric dilaton-gravity action in 2D
$$S^{(1)}={i\over 2\pi}\ixt\, E\left[
J(\Phi)S+iK(\Phi)D_\alpha \Phi D^\alpha \Phi + L(\Phi)
\right]
\eqn\qi$$
($\Phi$ is the dilaton superfield, $S$ the survature
multiplet and $E$ the super-zweibein, see ref. \PS\ for
all notation and conventions) the purely bosonic part
(all fermi fields set to zero) reads after integrating
out the auxiliary fields
$$S^{(1)}_{\rm bos}={1\over 2\pi}\ix\rg\left[ J\, R +
2K\,
(\N\f)^2 +\left({LL'\over 2J'}-{KL^2\over
2J'^2}\right)\right]\ ,
\eqn\qii$$
where now $J=J(\f), K=K(\f), L=L(\f)$.

We now repeat this exercise, including a supersymmetric
$Z$-field:
$$S^{(2)}={i\over 2\pi}\ixt E\, \left[
-{i\over 4}D_\alpha {\cal Z} D^\alpha {\cal Z} + Q{\cal
Z}S \right]\ .
\eqn\qiii$$
The bosonic part of this action alone is just $S_Z$ of
eq. \dxiv\ after integrating out the auxiliary fields.
When combining $S^{(1)}$ and $S^{(2)}$, the auxiliary
field equations get modified and the resulting bosonic
part is not just \qii\ plus $S_Z$, but rather
$$\left[S^{(1)}+S^{(2)}\right]_{\rm bos}
={1\over 2\pi}\ix\rg\left[ J\, R +
2K\,  (\N\f)^2 -\half (\N Z)^2+QRZ + F(\f)\right]
\eqn\qiv$$
where
$$F(\f)=\left(1-{2\k K\over J'^2}\right)^{-1}
\left({LL'\over 2J'}-{KL^2\over
2J'^2}-{\k L'^2\over 4 J'^2}\right)\ .
\eqn\qv$$
All we have to do now is to identify the functions $J, K$
and $L$ of $\f$ that reproduce e.g. the RST-action \dxxii.

Equation \qiv\ will be identical to the RST-action \dxxii\
if
$$J(\f)=e^{-2\f}-{\k\over 2}\f\ ,\quad
K(\f)=2e^{-2\f}\ ,\quad
F(\f)=4\l^2 e^{-2\f}\ .
\eqn\qvi$$
Substituting these into eq. \qv\ we obtain a non-linear
differential equation for $L(\f)$ :
$$(L+xL')(L+L')=-\k^2\l^2 {(1-x)^2\over x^2}
\eqn\qvii$$
where $L'={\rm d}L/{\rm d}\f$ and $x={\k\over 4}e^{2\f}$.
The solution is very simple:
$$L(\f)=\pm4\l\left( e^{-2\f} +{\k\over 4}\right) =
\mp 2 \l J'(\f)\ .
\eqn\qviii$$
Obviously there are two choices of sign since only $\l^2$
is relevant. Thus, if $J, K$ and $L$ are given by \qvi,
\qviii, the action $S^{(1)}+S^{(2)}$ is a supersymmetric
extension of the RST-action. Note that the $Z$-independent
part of the energy-functional $M$ is $\sim \eb\gf (J' \NS\f
\pm \half L)\e$ as expected (cf. eq. (63) of ref. \PS).
Thus it can be derived from the square of the supercharge.

Similarly, we can construct a supersymmetric extension of
the action of ref. \BC. In this case $J(\f)$ and $K(\f)$
are given by the CGHS-functions
$$J(\f)=e^{-2\f}\ , \quad K(\f)=2 e^{-2\f}
\eqn\qix$$
while the function $F$ is more complicated [\BC]:
$$\eqalign{
F(\f)&=4\l^2e^{-2\f}D(\f)\cr
&=\k\l^2\, {1\over y}\left( 1+\sqrt{1-y}\right)^2\exp\left[
{1-\sqrt{1-y}\over 1+\sqrt{1-y}}\right]
}
\eqn\qx$$
where now $y=\k e^{2\f}$. The differential equation for
$L(\f)$ then is
$$LL'+L^2+{y\over 4}L'^2=-4\k\, {1-y\over y} F\ .
\eqn\qxi$$
If we substitute
$$L(\f)=2\l\k
\left(
{1+\sqrt{1-y}\over 1-\sqrt{1-y}}\right)^{1/2}
\exp\left[{1\over 2}\,
{1-\sqrt{1-y}\over 1+\sqrt{1-y}}\right]\,  g(y)
\eqn\qxii$$
the differential equation for $L$ simplifies to
$$2 y\sqrt{1-y}\, gg' +y^3 g'^2 =-{1-y\over y}
\eqn\qxiii$$
where $g'={\rm d}g/{\rm d}y$. If we now change variables
from $y=\k e^{2\f}$ to $\O$ with $\O$ given by \diii, the
differential equation becomes simply
$$g\, {{\rm d}g\over {\rm d}\O} -{1\over 4}
\left({{\rm d}g\over {\rm d}\O}\right)^2 = 1\ .
\eqn\qxiv$$
This is easily integrated and the solution $g(\O)$ is given
implicitly by the following transcendental equation ($c$ is
a constant of integration)
$$4(\O+c)=g(g \mp \sqrt{g^2-1})\pm \log
(g+\sqrt{g^2-1})\ .
\eqn\qxv$$
This defines the solution $g$, and by \qxii\ also $L(\f)$,
as a transcendental function of $\O(\f)$. The first terms
in an expansion for small $\k e^{2\f}$ are
$$L(\f)\sim \pm 4\l \left( e^{-2\f} +{\k\over 4} \f +\tilde
c\right)\ .
\eqn\qxvi$$
Although we have not derived it above, the
energy-functional $M$ should be obtained from the square of
the supercharge. Following ref. \PS\ and our observation
above we expect that the $Z$-independent part of the
energy-functional $M$ for the variant of ref. \BC\ is again
$\sim \eb\gf (J' \NS\f \pm \half L)\e$ (although now $L(\f)$
as given by \qxii\ and \qxv\ is a rather complicated
function!), while the $Z$-dependent part should be the same
as given in section 3 for the RST theory.

We have explicitly shown that the exactly solvable
conformally invariant actions of refs. \RST\ and \BC\ have
supersymmetric extensions. At first sight this seems to be
in contrast with the statement of ref. \NOJ\ that such
supersymmetric extensions do not exist. A closer look
however shows that one is dealing with two different
requirements. This was recently clarified by Danielsson
[\DAN] after a first circulation of the present paper.
Indeed what we claim here is to have constructed
{\it (semi)classical} theories that are the supersymmetric
extensions of the exact conformal theories of refs \BC\ and
\RST. The point is [\DAN] that integrating out the
auxiliary fields is a procedure that can only be trusted
semiclassically if the auxiliary fields are not set equal
to zero by their field equations. The reason is very simple
to see in the present case: integrating out the auxiliary
fields will typically replace a vertex operator of
conformal dimension $(\a,\a)$ by its square which {\it
classically} has dimension $(2\a,2\a)$, but of course not
quantum mechanically. Thus if we start with dimension
$(\half,\half)$ operators as required by exact {\it
superconformal} invariance we will not get $(1,1)$
operators after integrating out the auxiliary fields, and
vice versa. Since we insisted here on having a $(1,1)$
operator after integrating out the auxiliary fields, we
certainly did not have an exact superconformal theory to
start with. The claim of ref. \NOJ\ was precisely that such
an exact superconformal theory with the required bosonic
part does not exist. However, it turned out [\DAN] that by
complicating the original supersymmetric action slightly
one can construct an exact superconformal theory ($c=0$)
whose bosonic part, although not identical to any of the
exact bosonic conformal theories, still gives the usual
dilaton gravity in the (weak-coupling) semiclassical limit.

{\chapter{Conclusions}}

We have proven a positive energy (mass) theorem for the
exactly solvable quantum-corrected 2D dilaton-gravity
theory {\it \`a la} RST. Although there are probably many
more or less reasonable mass functionals, the one given
here is (within a relatively broad class) the only one that
obeys a positivity theorem. For field configurations
asymptotic to the LDV we have shown that this mass
functional if defined on a space-like surface coincides with
the usual definition of the ADM-mass given by the
field asymptotics at $\s=+\infty$, plus a certain
(``quantum")-correction $-{\cal M}(\s=-\infty)$ depending,
via the $\e$-differential equation, on the fields on all of
$\S$. For light-like (null) $\S$, we have given a rather
detailed analysis of the resulting Bondi-mass and shown
that it exhibits all expected physical properties: besides
being non-negative, it equals the  ADM-mass in the far
past, is decreasing for $\k>0$ and constant for $\k=0$, and
gives the correct (positive) energy of the thunderpop in
the shock-wave scenario (up to the order we computed).

We also explicitly constructed supersymmetric extensions of
the exactly solvable theories of refs. \BC\ and \RST. The
squares of the supercharges give us the positive energy
functionals, as we could check for the RST variant.

\ack

It is a pleasure to acknowledge very stimulating
discussions with Curt Callan, Ian Kogan, Andy Strominger and
Larus Thorlacius.

\refout

\end